\documentclass[aps,prl,twocolumn]{revtex4}
\usepackage{amssymb}

\usepackage{graphicx}

\begin{document}

\title{Protected nodal electron pocket from multiple-${\bf Q}$ ordering in underdoped high temperature superconductors}
\author{N.~Harrison$^1$, S.~E.~Sebastian$^2$}
\affiliation{$^1$National High Magnetic 
Field Laboratory, Los Alamos National Laboratory, MS E536,
Los Alamos, New Mexico 87545\\
$^2$Cavendish Laboratory, Cambridge University, JJ Thomson Avenue, Cambridge CB3~OHE, U.K
}
\date{\today}

\begin{abstract}
A multiple wavevector (${\bf Q}$) reconstruction of the Fermi surface is shown to yield a profoundly different electronic structure to that characteristic of single wavevector reconstruction, despite their proximity in energy. We consider the specific case in which ordering is generated by ${\bf Q}_x=[2\pi a,0]$ and ${\bf Q}_y=[0,2\pi b]$ (in which $a=b=\frac{1}{4}$) $-$ similar to those identified in neutron diffraction and scanning tunneling microscopy experiments, and more generally show that an isolated pocket adjacent to the nodal point ${\bf k}_{\rm nodal}=[\pm\frac{\pi}{2},\pm\frac{\pi}{2}]$ is a protected feature of such a multiple-$\bf Q$ model, potentially corresponding to the nodal `Fermi arcs' observed in photoemission and the small size of the electronic heat capacity found in high magnetic fields $-$ importantly, containing electron carriers which can yield negative Hall and Seebeck coefficients observed in high magnetic fields. 
\end{abstract}

\pacs{PACS numbers: 71.18.+y, 71.45.Lr, 74.25.Jb, 75.40.Mg, 75.30.Fv}
\maketitle

It has been challenging to identify the origin of small Fermi surface pockets observed by quantum oscillations in underdoped YBa$_2$Cu$_3$O$_{6+x}$~\cite{doiron1, leboeuf1, audouard1, sebastian4, sebastian2,ramshaw1, sebastian1}, given the absence of a unique expectation for the electronic structure from theoretical predictions in this regime~\cite{norman1}. 
While the details of the ordering differs between models, all translational symmetry-breaking models proposed thus far involve a single translational vector ${\bf Q}$~\cite{chakravarty1,millis1} causing them to yield Fermi surface topologies with the same universal features $-$ hole pockets and/or open sheets at the nodal point ${\bf k}_{\rm nodal}=[\pm\frac{\pi}{2},\pm\frac{\pi}{2}]$ in the extended Brillouin zone and the possibility of an electron pocket at the antinodal point ${\bf k}_{\rm antinodal}=[\pi,0]$ \& $[0,\pi]$.
It becomes important to search for new models outside the realm of these universal features, which have difficulty simultaneously capturing quantum oscillations of the negative Hall and Seebeck coefficients in a magnetic field~\cite{leboeuf1,laliberte1} (interpreted in terms of an antinodal electron pocket) and the pseudogap at the same location in zero field observed by ARPES and scanning tunneling spectroscopy experiments~\cite{ding1,damascelli1, lee1,valla1,chang1,shen1, kanigel1, hossain1}. While it has been suggested that these models of translational symmetry breaking may only be relevant in the high magnetic field regime of quantum oscillations, experiments have yet to establish the effect of a magnetic field in changing the electronic structure~\cite{tranquada1,riggs1}.

\begin{figure}[htbp!]
\centering
\includegraphics[width=0.4\textwidth]{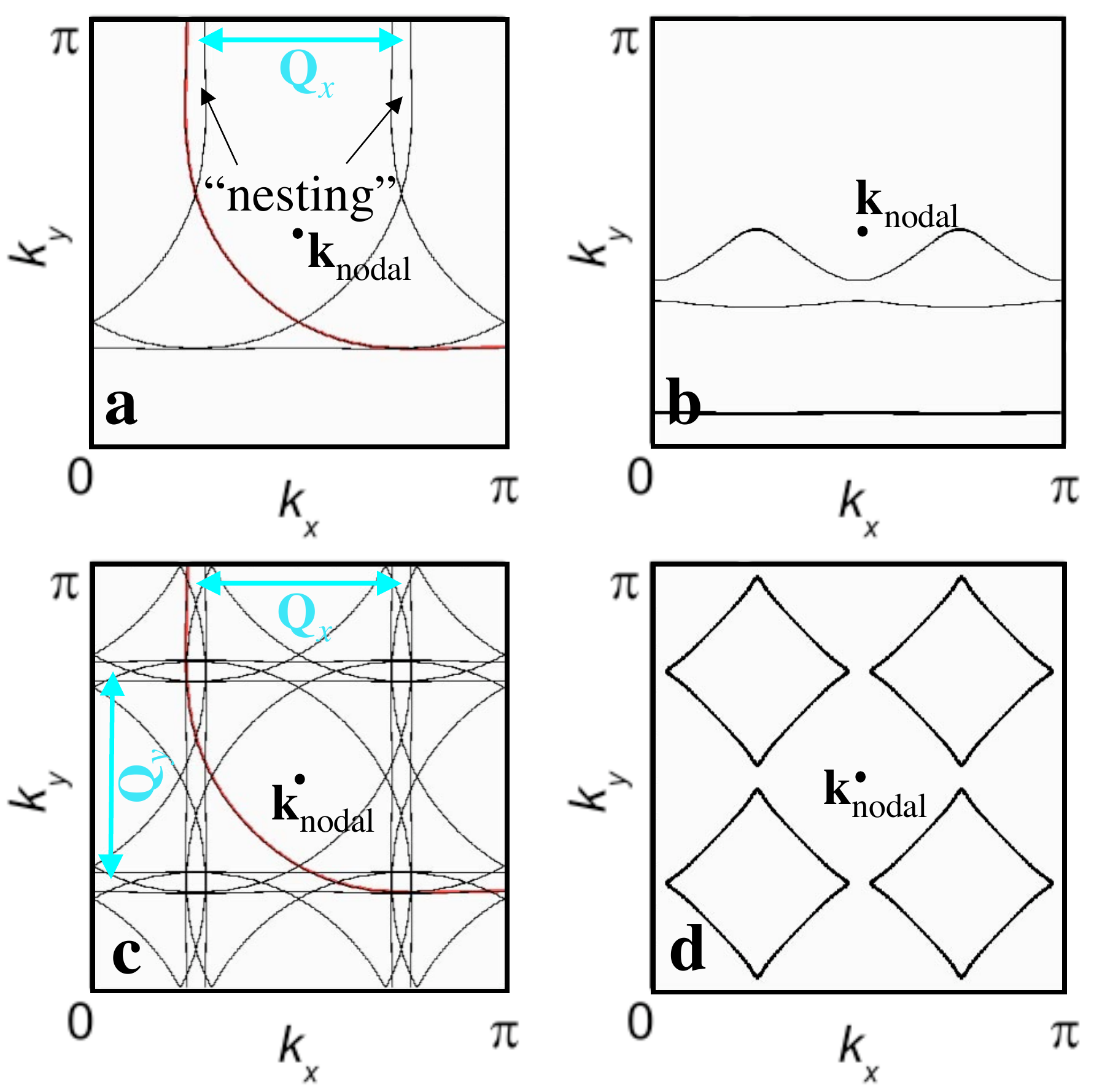}
\caption{{\bf a} Quadrant of the unreconstructed Fermi surface (red) together with its translation by multiples of ${\bf Q}_x$ (black) for $\delta=$~9~\%. {\bf b} Open sheets that result upon considering $V/W=0.1$. {\bf c} Same quadrant of the unreconstructed Brillouin zone together with its translation by multiples of ${\bf Q}_x$ and ${\bf Q}_y$. {\bf d} Example showing how $V/W=0.15$ leads to a diamond-shaped electron pocket.}
\label{single}
\end{figure}

Here we introduce a new possibility which we suggest is more applicable to the experimental situation in the underdoped cuprates. We show that multiple-${\bf Q}$ charge ordering yields a Fermi surface consisting of a protected electron pocket at the nodes $-$ rather surprisingly, while the multiple-${\bf Q}$ charge ordering solution is argued to be close in energy to the single-${\bf Q}$ charge ordering solution~\cite{melikyan1}, the consequences for electronic structure are profoundly different. 
In the case of underdoped YBa$_2$Cu$_3$O$_{6+x}$, a single electron pocket at the nodes produced by a range of coupling strengths in this model is most consistent with the single type of carrier pocket revealed by chemical potential quantum oscillations~\cite{sebastian5}, and the density-of-states concentrated chiefly at the nodes in ARPES~\cite{chang1,valla1,hossain1,kanigel1,shen1} and in-field heat capacity experiments~\cite{riggs1}; indeed the electron character of this pocket also yields a negative Hall and Seebeck coefficient as experimentally observed~\cite{leboeuf1,laliberte1}. We consider the case in which two orthogonal vectors ${\bf Q}_x=[2\pi a,0]$ and ${\bf Q}_y=[0,2\pi b]$ with $a\approx b\approx\frac{1}{4}$ lead to a $\approx$~16-fold reduction in the size of the Brillouin zone. Despite the aggregation of holes at ${\bf k}_{\rm nodal}$ in the extended Brillouin zone, their density exceeds 50~\% of the reconstructed Brillouin zone cross-section at the dopings relevant for quantum oscillation studies~\cite{doiron1, sebastian2} $-$ the electron pocket resulting from a $>$~50~\%~ filling of the folded band with holes. Unlike the antinodal electron pocket predicted by single-${\bf Q}$ models~\cite{millis1,chakravarty1}, the electron pocket found here incorporates the nodal `arc' region of the Fermi surface (see Fig.~\ref{single}) seen in ARPES~\cite{chang1,valla1,hossain1,kanigel1,shen1}.

Evidence for ordering at wavevectors ${\bf Q}_x$ and ${\bf Q}_y$ is found in neutron diffraction~\cite{tranquada2, fujita1, tranquada3} and scanning tunneling microscopy (STM)~\cite{hoffman1, hanaguri1, mcelroy1, wise1} measurements, with the possibilities for order including domains of unidirectional stripes incorporating spin order~\cite{robertson1}, stripes that alternate on consecutive layers~\cite{markiewicz1, zimmermann1, christensen1} or multiple-${\bf Q}$ charge ordering without a static spin component~\cite{note0,wise1, mcelroy1, shen1}. The latter may be relevant in underdoped YBa$_2$Cu$_3$O$_{6+x}$ where neutron scattering experiments find evidence for quasi-static spin ordering at oxygen compositions ($x<0.45$, corresponding to hole dopings $\delta\lesssim$~8~\%) lower than those in which quantum oscillations are observed~\cite{haug1}, while charge ordering is observed at the same and higher dopings (orthogonal~\cite{julien1} and parallel~\cite{liu1} to the chain direction respectively). Furthermore, the observation of spin zeroes in magnetic quantum oscillation experiments in strong magnetic fields is more easily explained by a scenario involving long range ordering of only the charge degrees of freedom~\cite{ramshaw1,sebastian1,norman2}.

We show in Fig.~\ref{single} how the multiple-${\bf Q}$ charge ordering model and single-${\bf Q}$ charge ordering scenarios, while close in energy~\cite{melikyan1}, lead to profoundly different electronic structures. If one considers the single-${\bf Q}$ Hamiltonian 
\begin{equation}\label{matrix0}
H_1=\left( \begin{array}{cccc}
\varepsilon & V & 0 & V\\
V & \varepsilon_{{\bf Q}_x} & V & 0\\
0 & V & \varepsilon_{2{\bf Q}_x} &V\\
V & 0 & V & \varepsilon_{3{\bf Q}_x}\\ \end{array} \right)
\end{equation}
for example, in which $V$ is the coupling and $\varepsilon_{{\bf Q}x}$ represents the electronic dispersion $\varepsilon({\bf k})$~\cite{bands,andersen1} translated by ${\bf Q}_x=[\frac{\pi}{2},0]$~\cite{millis1}, then susceptibility to gap formation (or approximate Fermi surface `nesting') is realized only near ${\bf k}=[0,\pi\pm\frac{\pi}{4}]$ in the extended Brillouin zone: not ${\bf k}=[\pi\pm\frac{\pi}{4},0]$ (see Fig.~\ref{single}{\bf a}). Diagonalization of $H_1$ yields open Fermi surface sheets (see Fig.~\ref{single}{\bf b}). By contrast, on considering simultaneous translations by ${\bf Q}_x$ and ${\bf Q}_y$, gap formation can occur both at ${\bf k}=[\pi\pm\frac{\pi}{4},0]$ and $[0,\pi\pm\frac{\pi}{4}]$ (see Fig.~\ref{single}{\bf c}). The open sheets of the single-${\bf Q}$ model give way to a small closed section of Fermi surface in Fig.~\ref{single}{\bf d} (on introducing a coupling $V$), necessary to obtain low frequency quantum oscillations~\cite{leboeuf1, doiron1, sebastian2}. 

\begin{figure*}[htbp!]
\centering
\includegraphics[width=0.7\textwidth]{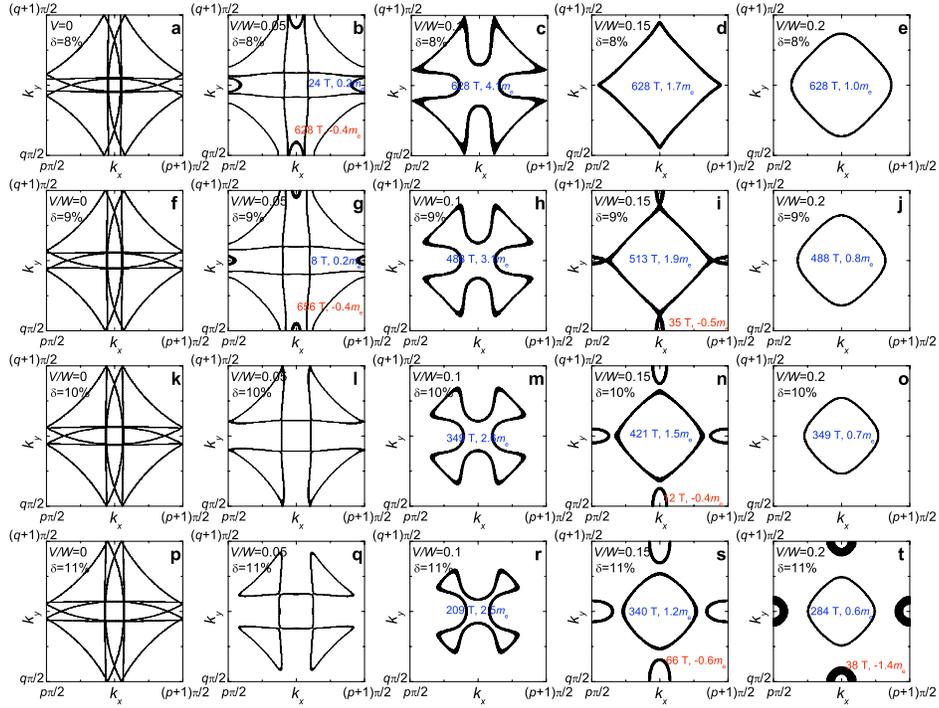}
\caption{Reconstructed Fermi surface according to Eqn (\ref{matrix}) for $a=b=\pm\frac{1}{4}$, different ratios $V/W$ and effective hole doping $\delta$ as indicated (assuming $V^\prime=V_{\rm II}=$~0). Also shown are the corresponding magnetic quantum oscillation frequencies according to Onsager's equation $F=A_k\hbar/2\pi e$ (where $A_k$ is the Fermi surface cross-section) and the corresponding band masses, where $m_{\rm e}$ is the free electron mass and a negative sign indicates holes. Line thicknesses are proportional to the inverse Fermi velocity.
}
\label{checkmate}
\end{figure*}

To construct a multiple-${\bf Q}$ ordering Hamiltonian, we consider all possible translations ${\bf k}\rightarrow{\bf k}+m{\bf Q}_x+n{\bf Q}_y$ in which $m$ and $n$ are integers to arrive at 
\begin{equation}\label{matrix}
H_2=\left( \begin{array}{cccccccc}
\varepsilon & V & V_{\rm II} & V & V  & V^\prime & \dots\\
V & \varepsilon_{{\bf Q}_x} & V & V_{\rm II}  & V^\prime  & V & \dots\\
V_{\rm II} & V & \varepsilon_{2{\bf Q}_x} &V & 0  & V^\prime  & \dots\\
V & V_{\rm II} & V & \varepsilon_{3{\bf Q}_x} & V^\prime & 0 &\dots\\
V & V^\prime & 0 & V^\prime & \varepsilon_{{\bf Q}_y}  &  V &\dots\\
V^\prime & V & V^\prime & 0 & V  &  \varepsilon_{{\bf Q}_y+{\bf Q}_x} &\dots\\
\vdots & \vdots & \vdots & \vdots & \vdots & \vdots & \ddots \end{array} \right).
\end{equation}
We initially neglect higher order terms (i.e. $V^\prime=V_{\rm II}=$~0), consider $V$ to be uniform and neglect a possible component of ${\bf Q}$ orthogonal to the layers $-$ similar Fermi surface results being envisaged~\cite{note1} for multiple-${\bf Q}$ d-density-wave order~\cite{soe1} or charge stripes that alternate on consecutive layers~\cite{markiewicz1, zimmermann1}. We further consider the commensurate case where $a=b=\pm\frac{1}{4}$ (relevant to many STM and neutron diffraction experiments), which yields a 16~$\times$~16 matrix and 16 bands upon diagonalization.  

Despite ${\bf k}_{\rm nodal}=[\pm\frac{\pi}{2},\pm\frac{\pi}{2}]$ having four equivalent locations in the extended Brillouin zone, multiple-${\bf Q}$ ordering folds these onto a single point in the reconstructed Brillouin zone yielding a concentration only of holes. The small size of the reconstructed Brillouin zone causes this hole surface area (corresponding to a frequency in the range 900~$<\frac{\hbar A_{\rm h}}{2\pi e}<$~1400~T, where $A_{\rm h}$ is the {\it k}-space area) to exceed 50~\%~of the zone area ($\frac{\hbar A_{\rm BZ}}{2\pi e}\approx$~1700~T) for nominal hole dopings (in the range 8~$\lesssim\delta\lesssim$~11~\%) applicable to quantum oscillation experiments~\cite{leboeuf1,doiron1,sebastian2} (see Fig.~\ref{single}{\bf d}). Band filling therefore causes an electron pocket to become a protected feature of multiple-${\bf Q}$ ordering. 

\begin{figure}[htbp!]
\centering
\includegraphics[width=0.45\textwidth]{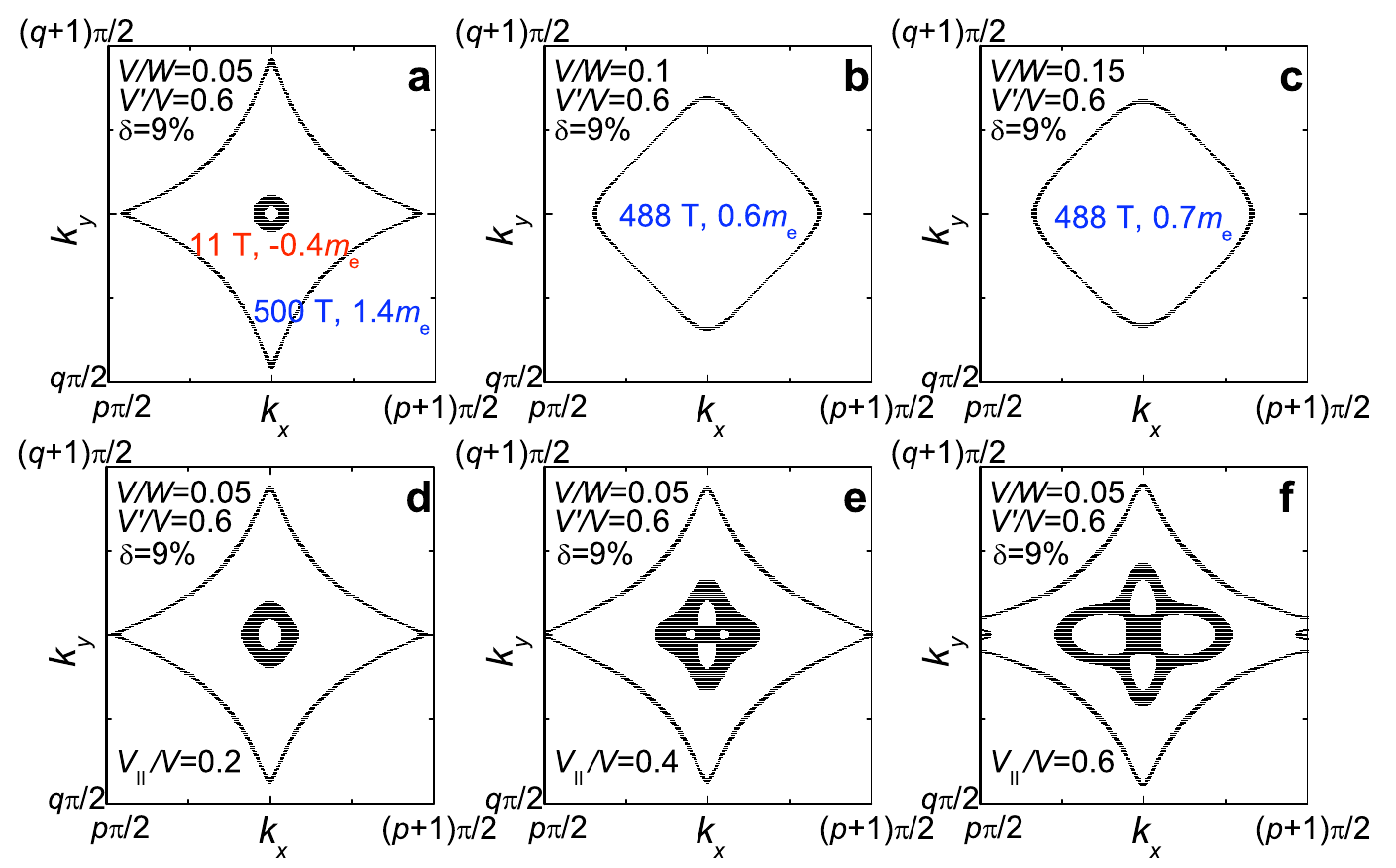}
\caption{{\bf a}, {\bf b} and {\bf c}. Reconstructed Fermi surface according to Eqn (\ref{matrix}) for $a=b=\pm\frac{1}{4}$ and different values of the ratio $V/W$ as indicated for an effective hole doping $\delta=$~9\% with $V^\prime/V=$~0.6 and $V_{\rm II}=$~0. {\bf d}, {\bf e} and {\bf f}. Same as ({\bf a}) but with the effect of different ortho-II potentials ($V_{\rm II}$) included.}
\label{simple}
\end{figure}

Fermi surfaces calculated for different combinations of $\delta$ and ratios $V/W$ of $V$ to the electronic bandwidth $W$~\cite{bands} are shown in Fig.~\ref{checkmate}. The complexity of the Fermi surface obtained at weaker couplings ($V/W<$~0.15) in Fig.~\ref{checkmate} originates from the combined effects of imperfect `nesting' and our neglect of the next nearest coupling $V^\prime$ between Fermi surfaces translated by ${\bf Q}_x\pm{\bf Q}_y$. These `nest' almost as well (see Fig.~\ref{single}{\bf c}) as those with a relative translation of ${\bf Q}_x$ or ${\bf Q}_y$. On including $V^\prime$ in Fig.~\ref{simple}{\bf a}-{\bf c}, a simple electron pocket is obtained for weaker coupling strengths. Finally, in Figs.~\ref{simple}{\bf d}-{\bf f} we include the effect of the ortho-II ordering potential $V_{\rm II}$, which breaks the rotational symmetry of the Fermi surface. For the case considered  ($V^\prime/V=$~0.6 and $V/W=$~0.05), the electron pocket remains intact provided $V_{\rm II}/V\leq$~0.4.

We compare features of the calculated Fermi surfaces in Fig.~\ref{checkmate} (for $V/W\sim$~0.15) and Fig.~\ref{simple} (for $V/W\sim$~0.05) with observed magnetic quantum oscillations in underdoped YBa$_2$Cu$_3$O$_{6+x}$ $-$ a single carrier type of pocket unaccompanied by a significant reservoir being indicated by the observation of chemical potential quantum oscillations~\cite{sebastian5}. Both the frequency $F\approx$~500~T and effective mass $m^\ast\approx$~1 - 2~$m_{\rm e}$ of the largest section obtained within the multiple-$\bf Q$ model we propose are close to those observed in quantum oscillation experiments~\cite{leboeuf1,doiron1,sebastian1,sebastian2,audouard1,sebastian3,sebastian4,ramshaw1,riggs1, laliberte1} $-$ no adjustment having been made to the tight-binding representation of the dispersion obtained from band structure calculations~\cite{bands,andersen1}. As opposed to previously proposed single-$\bf Q$ models in which two nodal hole pockets are contained in each bilayer Brillouin zone, the small size of the Brillouin zone in the multiple-$\bf Q$ model we propose here causes it to contain only a single dominant electron pocket, yielding values for the Sommerfeld coefficient  $\gamma_{\rm model}\approx$~5 - 9~mJmol$^{-1}$K$^{-2}$ (considering $V/W=$~0.15 and counting 2 bilayers) comparable to that $\gamma_{\rm exp}\approx$~5.3~mJmol$^{-1}$K$^{-2}$ obtained in heat capacity studies in strong magnetic fields~\cite{riggs1}. The upper end of this range is caused by proximity to a Lifshitz transition at $\delta\approx$~9~\% when $V^\prime=$~0 in the model. Furthermore, the electron-character of the largest predicted pocket over a broad range of dopings and couplings (being the only pocket for many combinations of $V$ and $\delta$) yields a negative Hall and Seebeck coefficients, as seen in strong magnetic fields in underdoped YBa$_2$Cu$_3$O$_{6+x}$~\cite{leboeuf1, laliberte1}.

\begin{figure}[htbp!]
\centering
\includegraphics[width=0.4\textwidth]{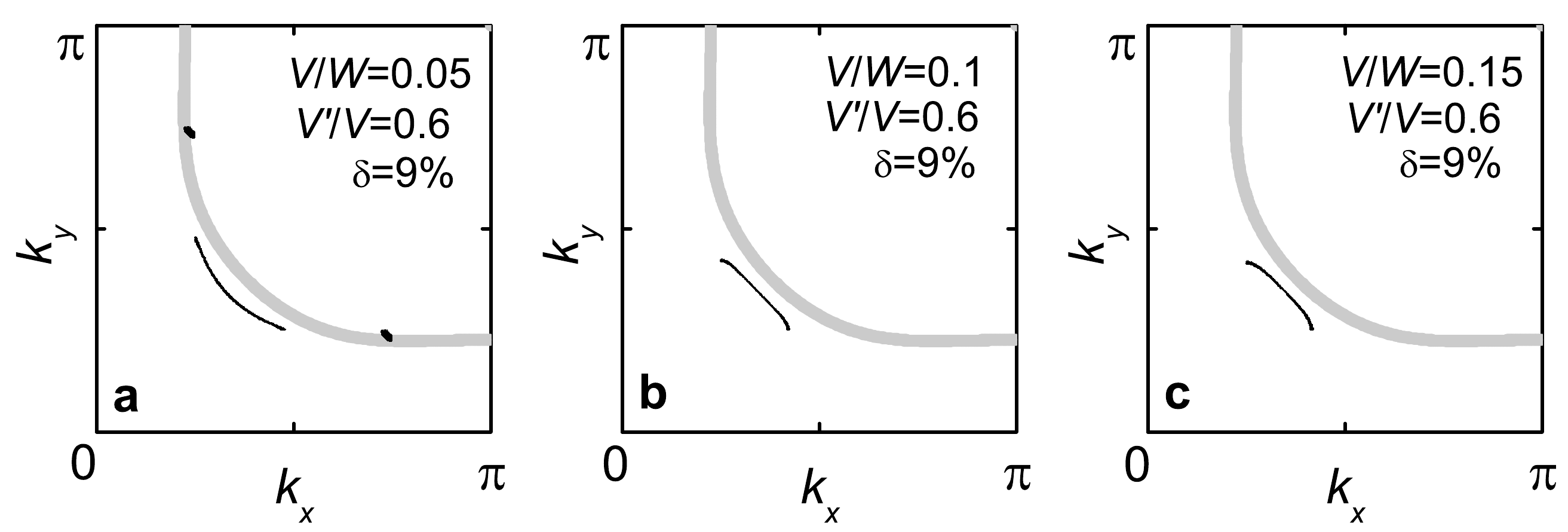}
\caption{The result of `unfolding' the reconstructed Fermi surfaces in Figs.~\ref{simple}{\bf a}-{\bf c} so as to trace the origin of the Fermi surface segments to the original Brillouin zone (only $1/4$ of which is shown).}
\label{arcs}
\end{figure}

Features of the multiple-$\bf Q$ model are also compared with ARPES (and STM) Fermi surface measurements. Since the predicted electron pocket is constructed entirely from the nodal regions of $\varepsilon({\bf k})$ and is also the largest pocket, the observation of Fermi arcs~\cite{shen1, kanigel1, hossain1, lee1} would not be unexpected in this scenario. Specifically, as a consequence of coherence factors~\cite{chakravarty2} and short correlation lengths~\cite{harrison2}, ARPES experiments may be expected to be sensitive chiefly to the portions of the reconstructed Fermi surface (black lines in Fig.~\ref{arcs}) that overlap with the unreconstructed Fermi surface (depicted in grey). 

The proximity of the electron pocket to the reconstructed Brillouin zone boundary (or to small hole pockets for certain values of $\delta$ and $V$ in Fig.~\ref{checkmate}) introduces the further possibility of magnetic breakdown~\cite{shoenberg1}. Multiple frequencies clustered around the prominent frequency at $F\approx$~500~T in experiments~\cite{audouard1} can also be caused by bilayer splitting, which if included would result in couplings both within and between bonding and antibonding bands in a charge ordering model~\cite{note2}. Given the small size of the reconstructed Brillouin zone in the model we consider here, a possible Brillouin zone frequency~\cite{shoenberg1,harrison1} may additionally arise, of size close to the reported $\beta$ frequency~\cite{sebastian4} ($\approx$~1650~T).

In summary, we identify an electron pocket located at the nodal regions of the Brillouin zone as a universal feature of multiple-$\bf Q$ charge ordering. A viable alternative is proposed to the previously proposed spin and/or charge models which lead to a combination of hole pockets at the nodes, electron pockets at the antinodes, and open sheets of Fermi surface. The applicability of this model to underdoped YBa$_2$Cu$_3$O$_{6+x}$ in which small Fermi surface pockets have been observed is particularly relevant given the recent discovery of charge order in the chains in this material at high magnetic fields~\cite{julien1} $-$ the size, location and carrier type of the Fermi surface topology expected within this model are shown to be better consistent with experimental observations than previously proposed single-$\bf Q$ models. A further strength of the multiple-${\bf Q}$ model is that that the electron pocket is formed from the lowest of the upper 8 bands in the multiple-${\bf Q}$ model (out of a total 16 bands), suggesting that this pocket will remain robust against the introduction of a strong Coulomb repulsion in the model.

This work is supported by DOE BES project ``Science at 100 Tesla.'' We acknowledge helpful advice from G.~G.~Lonzarich.


\begin{thebibliography}{99}
\bibitem{note2} Couplings between bonding and antibonding bands would yield two degenerate frequencies $F$ whereas coupling within bonding and antibonding bands would yield frequencies $F\pm\Delta F$ where $\Delta F$ depends on the magnitude of the bilayer splitting in the nodal region.
\end{thebibliography}
\end{document}